\documentclass[12pt]{article}
\usepackage{amsmath}
\usepackage{latexsym}

\newcommand{\out}{\operatorname{out}}
\newcommand{\In}{\operatorname{in}}

\begin{document}

\title{Vacuum Electrodynamics of Accelerated Systems: Nonlocal Maxwell's Equations}

\author{Bahram Mashhoon\\Department of Physics and Astronomy\\University of Missouri-Columbia\\Columbia,
Missouri 65211, USA\\ E-Mail: \texttt{MashhoonB@missouri.edu}}

\maketitle

\begin{abstract} The nonlocal electrodynamics of accelerated systems is discussed in connection with the
development of Lorentz-invariant nonlocal field equations. Nonlocal Maxwell's equations are presented
explicitly for certain linearly accelerated systems. In general, the field equations remain nonlocal even
after accelerated motion has ceased.\end{abstract}

\noindent\footnotesize{PACS numbers: 03.30.+p, 11.10.Lm}\medskip

\setlength{\baselineskip}{24pt}

\section{Introduction}\label{s1}

In the description of dynamics as well as electrodynamics of continuous media, one often comes across
memory-dependent phenomena (``after-effects"). The resulting nonlocal characterization of the properties
of continua has a long history (see,
e.g., \cite{1}-\cite{3}). According to Landau and Lifshitz~\cite{4}, the most general linear relationship
between the electric induction ${\bf D}$ and the electric field ${\bf E}$ is given by
\begin{equation}\label{eq1} {\bf D}(t)={\bf E}(t)+\int^t_{-\infty} f(t-\tau ){\bf E}(\tau )d\tau
,\end{equation} where $f$ depends on the properties of the medium. It
is assumed here that at any field point ${\bf x} = (x,y,z)$, ${\bf D}$
depends upon the value of ${\bf E}$ at all previous moments; thus, for
a static inertial observer at ${\bf x}$, the integration in
\eqref{eq1} extends over the past worldline of the observer. This Volterra integral equation of the
second kind is consistent with the requirement of causality and has a kernel of the form $K(t,\tau
)=f(t-\tau )$. Such an equation is often called an integral equation of the Poisson type, since, according
to \cite{5}, Poisson first considered kernels of convolution (Faltung) type in his theory of induced
magnetism~\cite{2,3}. Such kernels also appear in the treatment of hysteresis~\cite{6}. Volterra integral
equations with convolution kernels are thus very important for physical applications; in fact, such
equations---called ``equations of the closed cycle" by Volterra---were used in Volterra's theory of
heredity in mechanical systems~\cite{7}. As explained by Tricomi~\cite{8}, the closed-cycle designation
originates from the fact that the operation
\begin{equation}\label{eq2} \Phi_{\out} (x)=\int^x_{-\infty}K(x,y)\Phi_{\In} (y)dy\end{equation} carries
a periodic input function $\Phi_{\In }$ with period $T$ into a periodic output function $\Phi_{\out}$
with {\it the same period} $T$ if and only if the kernel $K(x,y)$ is of the convolution type.

Nonlocal constitutive relations such as~\eqref{eq1} naturally lead to nonlocal Maxwell's equations for
the electrodynamics of continuous media~\cite{9}. In this paper, we are interested in nonlocal Maxwell's
equations in the absence of any medium; indeed, the nonlocality is induced by the fact that
electromagnetic measurements are performed by {\it accelerated observers}. That is, the nonlocality under
consideration in this paper is a characteristic of the vacuum state as perceived by accelerated
observers~\cite{10}.

Acceleration-induced nonlocality is described in Section~\ref{s2}. Section~\ref{s3} contains certain
mathematical results that are needed in order to derive nonlocal Maxwell's equations in Section~\ref{s4}.
An alternative treatment of the nonlocal field equations based on the vector potential is presented in
Section~\ref{s5}. Section~\ref{s6} contains a brief discussion of our results.

\section{Acceleration-induced nonlocality}\label{s2}

Consider a noninertial observer following a path $x^\mu (\tau )$ in Minkowski spacetime. Here $x^\mu
=(ct,x,y,z)$ are global inertial coordinates and $\tau$ is the proper time of the observer such that along
the worldline $-d\tau^2=\eta_{\mu
\nu}dx^\mu dx^\nu$, where $\eta_{\mu \nu}$ is the Minkowski metric tensor with signature $+2$. According
to the standard theory of special relativity, a noninertial observer at each instant along its worldline is
equivalent to an otherwise identical momentarily comoving inertial observer. In effect, the noninertial
observer may be replaced by a continuous infinity of hypothetical momentarily comoving inertial observers
along its worldline. This {\it hypothesis of locality} is the basis for the extension of Lorentz
invariance to noninertial observers~\cite{11}-\cite{13}.

It follows from the hypothesis of locality that each noninertial observer is endowed with an orthonormal
tetrad frame
$\lambda ^\mu_{(\alpha)}(\tau )$ such that $\lambda^\mu _{(0)}=dx^\mu /d\tau$ constitutes the temporal
axis of the observer's local frame and $\lambda^\mu_{(i)}$, $i=1,2,3$, are unit spatial directions that
constitute the local spatial frame of the observer. Along the observer's worldline
\begin{equation}\label{eq3} \frac{d\lambda^\mu _{(\alpha)}}{d\tau
}=\phi_\alpha ^{\;\;\;\beta} \; \lambda^\mu
_{(\beta )},\end{equation} where $\phi_{\alpha \beta}$ is the antisymmetric acceleration tensor. In
analogy with the Faraday tensor, we can write
$\phi_{\alpha \beta}\to (-{\bf a}$ , \boldmath{$\Omega$}\unboldmath $)$, where ${\bf a}(\tau )$ is the
translational acceleration $(\phi_{0i}=a_i)$ and \boldmath$\Omega$\unboldmath$(\tau )$ is the frequency
of rotation of the local spatial frame with respect to a nonrotating---that is, Fermi-Walker
transported---frame $(\phi_{ij}=\epsilon_{ijk}\Omega^k)$.

The scalar quantities $\phi_{\alpha \beta}$ lead to the existence of invariant acceleration scales, i.e.
the lengths
$\mathcal{L}=c^2/a$ and $c/\Omega$ and the corresponding acceleration times $c/a$ and $1/\Omega$,
respectively. These characterize the scale of variation of the state of the observer given at each instant
by its position and velocity. If the intrinsic scale of the phenomenon under observation is negligible
compared to the observer's acceleration scale, then the hypothesis of locality is a valid approximation.
This holds for most laboratory phenomena, since Earth-based observers have
$c^2/g_\oplus \simeq 1$ lyr and $c/\Omega _{\oplus} \simeq 28$ au. The deviation from the hypothesis of
locality is expected to be proportional to $\lambda/\mathcal{L}$, where $\lambda$ is the characteristic
length scale of the phenomenon under observation. Consider, for instance, the reception of an
electromagnetic wave of wavelength $\lambda$ by the observer. A few periods of the wave must be
registered before the observer can determine the properties of the incident radiation such as its
frequency; meanwhile, the observer's velocity changes. If this change is negligibly small over the period
of observation, then $\lambda /\mathcal{L} <<1$ and the assumption of locality is a reasonable
approximation. The hypothesis of locality holds exactly only for {\it coincidences}, i.e. once all
phenomena are reduced to the instantaneous interaction of classical point particles and rays of radiation.

To go beyond the hypothesis of locality, a nonlocal theory of accelerated observers has been developed
that takes the past history of the observer into account~\cite{14}. Consider, for instance, an
electromagnetic radiation field in the global inertial frame given by the Faraday tensor $F_{\mu \nu}
(x)$. Along the worldline of the accelerated observer, the class of momentarily comoving inertial
observers measures
\begin{equation}\label{eq4}\hat{F}_{\alpha \beta}(\tau )=F_{\mu \nu}\lambda^\mu _{(\alpha )}\lambda^\nu
_{(\beta)},\end{equation} which is the projection of the Faraday tensor onto the tetrad of the observer.
Let
$\hat{\mathcal{F}}_{\alpha \beta}$ be the Faraday tensor measured by the noninertial observer. The most
general linear relationship between
$\hat{\mathcal{F}}_{\alpha\beta}$ and $\hat{F}_{\alpha \beta}$ that is consistent with causality is
\begin{equation}\label{eq5} \hat{\mathcal{F}}_{\alpha \beta }(\tau )=\hat{F}_{\alpha \beta} (\tau
)+\int^\tau _{\tau_0}K_{\alpha
\beta}^{\;\;\;\;\gamma \delta} (\tau ,\tau ')\hat{F}_{\gamma \delta} (\tau ')d\tau ',\end{equation} where
$\tau_0$ is the instant at which the acceleration of the observer begins and $K$ is a kernel that is
expected to be proportional to the acceleration of the observer. The nonlocality of field determination
postulated in this ansatz is consistent with the ideas put forward by Bohr and Rosenfeld~\cite{15,16}; in
fact,~\eqref{eq5} involves a weighted average over the past worldline of the observer.

The Volterra integral relation~\eqref{eq5} implies that along the worldline, the relationship between
$\hat{\mathcal{F}}_{\alpha \beta}$ and $F_{\mu \nu}$ is unique in the space of continuous
functions~\cite{7}. Volterra's uniqueness theorem has been extended to the Hilbert space of
square-integrable functions by Tricomi~\cite{8}.

It is interesting to express equation~\eqref{eq5} in matrix form. To this end, let us write~\eqref{eq4}
as $\hat{F}=\Lambda F$, where $F$ and $\hat{F}$ are column vectors and $\Lambda$ is a $6\times 6$ matrix.
Here we employ the natural decomposition
$F_{\mu \nu}\to ({\bf  E},{\bf B})$ and then form the $6$-vector $F$ from the components of ${\bf E}$ and
${\bf B}$, respectively. The matrix $\Lambda$ is a representation of the Lorentz group. More generally,
let $x'=Lx+a$ be a passive proper Poincar\'e transformation from the global inertial frame to the
``local" inertial frame of the hypothetical momentarily comoving inertial observer; then, $\psi
'(x')=\Lambda (L)\psi (x)$, where $\psi$ is the field under consideration. The events along the worldline
are characterized by the proper time $\tau$; hence, $\hat{\psi }(\tau )=\Lambda (\tau )\psi (\tau )$ is
the field measured by the comoving inertial observer. For electromagnetism, \eqref{eq5} can be written as
\begin{equation}\label{eq6} \hat{\mathcal{F}}(\tau )=\Lambda (\tau )F(\tau )+\int^\tau _{\tau_0}K(\tau
,\tau ')\Lambda (\tau ')F(\tau ')d\tau ',\end{equation} where $K$ is a $6\times 6$ matrix.

In order to determine the kernel $K$, we postulate an extension of a well-known consequence of Lorentz
invariance to all observers, namely, that a basic radiation field can never stand completely still with
respect to any observer. For inertial observers, the Doppler effect relates the frequency measured by an
observer moving with uniform velocity ${\bf v}$ with respect to static observers in an inertial frame,
i.e. $\omega '=\gamma (\omega -{\bf v}\cdot {\bf k})$, where $\omega$ and
${\bf k}$ are the frequency and wave vector of the radiation as measured by the inertial observers such
that $\omega =c|{\bf k}|$. We note that $\omega '=0$ if and only if $\omega =0$; therefore, we require
that if $\hat{\mathcal{F}}_{\alpha
\beta}(\tau )$ turns out to be a constant, then $F_{\mu \nu}(\tau )$ must have been a constant in the
first place. In that case, the Volterra-Tricomi uniqueness theorem guarantees that for a true time-dependent
radiation field $F_{\mu \nu }(\tau )$, the field measured by the accelerated observer
$\hat{\mathcal{F}}_{\alpha \beta}(\tau )$ will be time-dependent so that a true radiation field would
never stand completely still with respect to any observer. It follows from~\eqref{eq6} that
$\hat{\mathcal{F}}(\tau _0)=\Lambda (\tau _0)F(\tau _0)$; therefore, our postulate implies that
\begin{equation}\label{eq7}\Lambda (\tau _0)=\Lambda (\tau )+\int^\tau_{\tau_0}K(\tau ,\tau ')\Lambda
(\tau ')d\tau '.\end{equation} Given $\Lambda (\tau )$, this equation may be used to determine the kernel
$K$. However, \eqref{eq7} is not sufficient to determine $K$ uniquely. One possibility would be to assume
that $K(\tau ,\tau ')$ depends only on $\tau -\tau '$ as in the nonlocal electrodynamics of media
discussed in Section~\ref{s1}. In fact, this assumption was made in \cite{14,17} as it appeared to be the
most natural one from a physical point of view. A detailed examination has revealed, however, that such a
convolution kernel leads to divergences in the case of non-uniform acceleration~\cite{18}. Further
examination of the solutions of~\eqref{eq7} together with a comparison of the results with experimental
data on spin-rotation coupling has revealed that the only acceptable solution of~\eqref{eq7} is \cite{19}
\begin{equation}\label{eq8}K(\tau ,\tau ')=k(\tau ')=-\frac{d\Lambda (\tau ')}{d\tau '}\Lambda ^{-1}(\tau
').\end{equation} It is interesting to note that in the case of
uniform acceleration this kernel is constant and
identical with the convolution kernel~\cite{14,17}. 

The unique kernel \eqref{eq8} has the property that it vanishes once the acceleration is turned off at $\tau _f$. For
$\tau >\tau _f$, the nonlocal part of equation~\eqref{eq5} is simply a constant field. This constant
memory of past acceleration is cumulative and hence could possibly become large enough to be detectable
at some time. On the other hand, the memory of past acceleration is canceled whenever a measuring device
is reset.

The observational consequences of this nonlocal electrodynamics of accelerated systems have been briefly
described in
\cite{20}. The present paper is devoted to the development of nonlocal field equations in this theory.

\section{Nonlocal electrodynamics}\label{s3}

Let us assume the existence of a nonlocal field $\mathcal{F}_{\mu \nu}(x)\to (\mathcal{E},\mathcal{B})$
such that an accelerated observer with tetrad frame $\lambda^\mu _{(\alpha )}$ would measure the
electromagnetic field
\begin{equation}\label{eq9} \hat{\mathcal{F}}_{\alpha \beta} =\mathcal{F}_{\mu \nu}\lambda^\mu _{(\alpha
)}\lambda^\nu_{(\beta)}\end{equation} given by
\begin{equation}\label{eq10} \hat{\mathcal{F}}=\hat{F}+\int^\tau_{\tau_0}k(\tau ')\hat{F} (\tau ')d\tau
',\end{equation} where $k(\tau ')$ is given by~\eqref{eq8}. It follows from the integral
equation~\eqref{eq10} that
\begin{equation}\label{eq11} \hat{F}=\hat{\mathcal{F}}+\int^\tau_{\tau_0}r(\tau ,\tau
')\hat{\mathcal{F}}(\tau ')d\tau ',\end{equation} where $r(\tau ,\tau ')$ is the resolvent kernel.
Equation~\eqref{eq11} may be re-expressed as
\begin{equation}\label{eq12} F=\mathcal{F}+\int^\tau _{\tau _0}\tilde{r}(\tau ,\tau ')\mathcal{F}(\tau
')d\tau ',\end{equation} where $\tilde{r}$ is related to the resolvent kernel $r$ by
\begin{equation}\label{eq13}\tilde{r} (\tau ,\tau ')=\Lambda^{-1}(\tau )\;\;r(\tau ,\tau ')\;\;\Lambda
(\tau ').\end{equation} The resolvent kernels $r$ and $\tilde{r}$ are discussed in detail in
Appendices~A-D.

Let us now imagine the possibility of extending~\eqref{eq12} to a class of accelerated observers. In this
way, the field
$F_{\mu \nu}(x)$ throughout a region of spacetime is related to a nonlocal field $\mathcal{F}_{\mu
\nu}(x)$ via the extended form of~\eqref{eq12}. The radiation field $F_{\mu \nu}(x)$ satisfies Maxwell's
source-free equations in Minkowski spacetime; therefore, one may derive the corresponding nonlocal field
equations for
$\mathcal{F}_{\mu\nu}(x)$. What any individual accelerated observer measures is then given
by~\eqref{eq9}, just as the field measured by an inertial observer would be the projection of $F_{\mu
\nu}$ onto the tetrad frame of the inertial observer. That is, the fundamental distinction between inertial
and accelerated observers is now reflected in the difference between the local field $F_{\mu \nu}$ and the
nonlocal field
$\mathcal{F}_{\mu \nu}$. This is illustrated in the next section for a special class of noninertial
observers. In fact, we consider the case of linearly accelerated observers for the sake of simplicity.

\section{Nonlocal Maxwell's equations}\label{s4}

Consider an observer that is at rest in the global background inertial frame for $-\infty <t<0$ and at
$t=0$ accelerates from rest with acceleration $g(\tau )>0$ along the positive $z$-direction. Here $\tau $
is the proper time along the observer's path such that $\tau =0$ at $t=0$. For $t\geq 0$, the nonrotating
orthonormal tetrad of the observer is given by
\begin{equation}\begin{split} \lambda^\mu_{(0)} & =(C,0,0,S),\\
\lambda^\mu_{(1)}&=(0,1,0,0),\\
\lambda^\mu_{(2)}&=(0,0,1,0),\\
\lambda^\mu_{(3)}&=(S,0,0,C),\label{eq14}\end{split}\end{equation} where $C=\cosh \theta$, $S=\sinh
\theta$ and
\begin{equation}\label{eq15} \theta=\frac{1}{c^2}\int^\tau_0g(\tau ')d\tau '.\end{equation} One can show
that in this case
\begin{equation}\label{eq16} \Lambda=\begin{bmatrix} U & V\\ -V & U\end{bmatrix},\quad U=\begin{bmatrix}
C & 0 & 0\\ 0 & C & 0\\ 0 & 0 & 1\end{bmatrix} ,\quad V=SI_3,\end{equation} where $I_i$,
$(I_i)_{jk}=-\epsilon_{ijk}$ is a $3\times 3$ matrix proportional to the operator of infinitesimal
rotations about the $x^i$-axis. Moreover, $UV=VU=CSI_3$ and $U^2+V^2=I$, where $I$ is the unit matrix.
These relations imply that
$\Lambda^{-1}(\tau )$ has the same form as $\Lambda (\tau )$ given in~\eqref{eq16} except for $V\to -V$.
It then follows from equation~\eqref{eq8} that
\begin{equation}\label{eq17} k(\tau )=-g(\tau )\begin{bmatrix} 0 & I_3\\-I_3 &
0\end{bmatrix},\end{equation} where henceforth units are chosen such that $c=1$.

We need to compute the resolvent kernel corresponding to~\eqref{eq17} and determine $\tilde{r}$ defined
by~\eqref{eq13}. A detailed calculation (in Appendix D) reveals that
\begin{equation}\label{eq18} \tilde{r}(\tau ,\tau')=-k(\tau '),\end{equation} so that~\eqref{eq12} takes
the form
\begin{equation}\label{eq19} F(\tau )=\mathcal{F}(\tau )+\begin{bmatrix} 0 & I_3\\ -I_3 & 0\end{bmatrix}
u(\tau )\int^\tau_0g(\tau ')\mathcal{F}(\tau ')d\tau ',\end{equation} where $u(\tau )$ is the unit step
function such that $u(\tau )=1$ for $\tau >0$ and $u(\tau )=0$ for $\tau <0$. In~\eqref{eq19}, the
nonlocal part has been multiplied by $u(\tau )$ to ensure that~\eqref{eq19} is valid for
all time $\tau :-\infty \to \infty$ along the worldline of the accelerated observer.

For the sake of simplicity, we now assume that $g(\tau )$ is uniform and equal to $g_0$ for $0<\tau
<\tau_f$ and zero otherwise, i.e.
\begin{equation}\label{eq20} g(\tau )=g_0\;\; [u(\tau )-u(\tau -\tau_f)].\end{equation} To extend the
validity of~\eqref{eq19} to a whole class of noninertial observers, consider the class of fundamental
static inertial observers in the background global inertial frame. Let us imagine that at $t=0$ the whole
class is accelerated uniformly with constant acceleration $g_0$ along the positive $z$-direction until
$t=t_f$, when the acceleration is turned off and the observers then move uniformly from $t_f$ on. For
$0\leq t\leq t_f$, the motion from $(0,x_0,y_0,z_0)\to (t,x,y,z)$ is given by $x=x_0$, $y=y_0$,
\begin{equation}\label{eq21} z=z_0+\frac{1}{g_0}(-1+\cosh g_0\tau ),\quad t=\frac{1}{g_0}\sinh g_0 \tau
,\end{equation} where $\tau =\tau _f$ at $t=t_f$. For $t>t_f$, the observers move with uniform speed
$\beta_f=\tanh g_0\tau _f$. The integrand in~\eqref{eq19} is in this case nonzero only during hyperbolic
motion; therefore, for $t>t_f$ the nonlocal part of~\eqref{eq19} for each observer simply involves a
constant field that is the memory of the observer's past acceleration. It proves useful to define the
function
\begin{equation}\label{eq22} \zeta (t)=\sqrt{t^2+\frac{1}{g_0^2}},\end{equation} so that during
hyperbolic motion $z(t)=z_0-g^{-1}_0+\zeta (t)$. Moreover, we define
\begin{equation}\label{eq23} \mathcal{U} (t)=u(t)-u(t-t_f).\end{equation} Thus~\eqref{eq19} can be
expressed as
\begin{equation}\label{eq24} F(t,x,y,z)=\mathcal{F}(t,x,y,z)+u(t)\begin{bmatrix} 0 & I_3\\ -I_3 &
0\end{bmatrix}
\int^t_0\mathcal{F} (t',x,y,z')\frac{\mathcal{U}'dt'}{\zeta '},\end{equation} where
$\mathcal{U}'=\mathcal{U} (t')$, $\zeta '=\zeta (t')$ and $z'=z_0-g^{-1}_0+\zeta '$. Eliminating the
initial position of the observer in favor of the field point ${\bf x}=(x,y,z)$, we finally have
\begin{equation}\label{eq25} F(t,{\bf x})=\mathcal{F}(t,{\bf x})+u(t)\begin{bmatrix} 0 & I_3\\ -I_3 &
0\end{bmatrix}
\int^t_0\mathcal{F}(t',x,y,z-\zeta +\zeta ')\frac{\mathcal{U}'dt'}{\zeta '}.\end{equation}

The substitution of $F(t,{\bf x})$ in Maxwell's equations would then result in nonlocal Maxwell's
equations for
$\mathcal{F}(t,{\bf x})$. To this end, let us note that~\eqref{eq25} can be written as
\begin{align}\label{eq26}{\bf E}&=\text{\boldmath$\mathcal{E}$\unboldmath} +u(t) \;\;\hat{{\bf z}}\times
\int^t_0\text{\boldmath$\mathcal{B}$\unboldmath} (t',x,y,z-\zeta +\zeta ')\frac{\mathcal{U}'dt'}{\zeta
'},\\
\label{eq27}{\bf B} &= \text{\boldmath$\mathcal{B}$\unboldmath} -u(t)\;\;\hat{{\bf z}} \times
\int^t_0\text{\boldmath$\mathcal{E}$\unboldmath}(t',x,y,z-\zeta +\zeta ')\frac{\mathcal{U}'dt'}{\zeta
'},\end{align} so that the fields remain unchanged along the direction of motion of the observer just as
under Lorentz transformations. Maxwell's equations with respect to inertial observers may be expressed as
\begin{align}\label{eq28} \text{\boldmath$\nabla$\unboldmath} \cdot {\bf W}^\pm &=0,\\
\label{eq29} \frac{1}{i}\text{\boldmath$\nabla$\unboldmath} \times {\bf W}^\pm &=\pm
\frac{\partial}{\partial t}{\bf W}^\pm,\end{align} where ${\bf W}^\pm ={\bf E}\pm i{\bf B}$ are the
Kramers vectors. In terms of complex field amplitudes, ${\bf W}^+({\bf W}^-)$ represents the positive
(negative) helicity state of the radiation field. Let us note that if~\eqref{eq28} is valid at some
initial instant of time, then this initial condition is maintained by~\eqref{eq29}, which is of the Dirac
form. Moreover,~\eqref{eq28} and~\eqref{eq29} imply that $\Box {\bf W}^\pm =0$. Defining
\boldmath$\mathcal{W}$\unboldmath$^\pm=$\boldmath$\mathcal{E}$\unboldmath
 $\pm\;  i$\boldmath$\mathcal{B}$\unboldmath, we note from~\eqref{eq26} and~\eqref{eq27} that
\begin{equation}\label{eq30} {\bf W}^\pm =\text{\boldmath$\mathcal{W}$\unboldmath}^\pm \mp
iu(t)\;\;\hat{{\bf z}}\times
\int^t_0\text{\boldmath$\mathcal{W}$\unboldmath} ^\pm (t',x,y,z-\zeta +\zeta
')\frac{\mathcal{U}'dt'}{\zeta '}.\end{equation} It then follows from~\eqref{eq28} and~\eqref{eq29} that
the nonlocal Maxwell equations are given by
\begin{align}\label{eq31} &\text{\boldmath$\nabla$\unboldmath}\cdot
\text{\boldmath$\mathcal{W}$\unboldmath}^\pm \pm iu(t)\;\;\hat{{\bf z}} \cdot
\int^t_0\text{\boldmath$\nabla$\unboldmath} \times \text{\boldmath$\mathcal{W}$\unboldmath}^\pm
(t',x,y,z-\zeta +\zeta ')\frac{\mathcal{U}'dt'}{\zeta '}=0,\\
 &\frac{1}{i} \left( \text{\boldmath$\nabla$\unboldmath}
-\frac{\mathcal{U}}{\zeta}\hat{{\bf z}}\right) \times \text{\boldmath$\mathcal{W}$\unboldmath}^\pm =\pm
\frac{\partial
\text{\boldmath$\mathcal{W}$\unboldmath}^\pm}{\partial t}\pm
u(t)\int^t_0\text{\boldmath$\nabla$\unboldmath}
\times (\hat{{\bf z}}\times
\text{\boldmath$\mathcal{W}$\unboldmath}^\pm )(t',x,y,z-\zeta+\zeta ')\frac{\mathcal{U}'dt'}{\zeta
'}\notag\\ &\quad +i \frac{tu(t)}{\zeta }\;\; \hat{{\bf z}} \times \int^t_0
\partial_z\text{\boldmath$\mathcal{W}$\unboldmath}^\pm (t',x,y,z-\zeta +\zeta
')\frac{\mathcal{U}'dt'}{\zeta '}.\end{align}

An important physical consequence of these equations is that \boldmath$\mathcal{W}$\unboldmath$^+$ and
\boldmath$\mathcal{W}$\unboldmath$^-$ satisfy separate equations, i.e. nonlocality cannot turn one
helicity state into another. Furthermore, the equations remain nonlocal for $t>t_f$. This significant
property originates from the fact that although each individual observer after $t_f$ measures a constant
additional field as its memory of past acceleration, these constant fields are different for different
observers. Thus the field equations remain nonlocal even after the acceleration has been turned off.

\section{Vector potential}\label{s5}

An alternative treatment of the electromagnetic equations would involve the vector potential $A_\mu (x)$
as the basic field, where $F_{\mu\nu}=A_{\nu,\mu}-A_{\mu,\nu}$, even though $A_\mu$ is only determined up
to a gauge transformation. The vector potential is thus devoid of direct physical significance in
classical electrodynamics; however, this is not the case in the quantum theory.

According to the hypothesis of locality
\begin{equation}\label{eq33} \hat{A}_\alpha =A_\mu \lambda^\mu _{(\alpha)}\end{equation} is the vector
potential measured by an accelerated observer, while the nonlocal theory asserts that this field is given
by
\begin{equation}\label{eq34} \hat{\mathcal{A}}_\alpha (\tau )=\hat{A}_\alpha (\tau )+u(\tau -\tau
_0)\int^\tau _{\tau _0}k_\alpha^{\;\;\beta} (\tau ')\hat{A}_\beta (\tau ')d\tau ',\end{equation} where
$\tau _0$ is the instant at which the acceleration is turned on. The kernel is given by
equation~\eqref{eq8} and we find that in general
\begin{equation}\label{eq35} k_{\alpha \beta}(\tau )=-\phi_{\alpha \beta}(\tau ),\end{equation} where
$\phi_{\alpha \beta}$ is the acceleration tensor defined by~\eqref{eq3}.

For the linearly accelerated observer with tetrad~\eqref{eq14},~\eqref{eq35} implies that $k=(k_\alpha
^{\;\;\beta})$ is given by
\begin{equation}\label{eq36} k(\tau )=-g(\tau )\begin{bmatrix} 0 & 0 & 0 & 1\\ 0 & 0 & 0 & 0\\ 0 & 0 & 0 &
0\\ 1 & 0 & 0 & 0\end{bmatrix}.\end{equation} It then follows from a detailed calculation that
\begin{equation}\label{eq37} r(x,y)=g(y)\begin{bmatrix} \mathcal{S} & 0 & 0 & \mathcal{C}\\ 0 & 0 & 0 &
0\\ 0& 0 & 0 & 0\\
\mathcal{C} & 0 & 0 &\mathcal{S}\end{bmatrix},\end{equation} where $\mathcal{C}=\cosh \sigma$,
$\mathcal{S}=\sinh \sigma$ and $\sigma =\theta (x)-\theta (y)$. Moreover,
$\tilde{r}(x,y)=-k(y)$, as expected. Thus,
\begin{align}\label{eq38} A_0(\tau )&=\mathcal{A}_0 (\tau )+u(\tau -\tau_0)\int^\tau_{\tau _0}g(\tau
')\mathcal{A}_3 (\tau ')d\tau ',\\
\label{eq39} A_1(\tau )&=\mathcal{A}_1(\tau ),\quad A_2(\tau )=\mathcal{A}_2(\tau ),\\
\label{eq40} A_3 (\tau )&=\mathcal{A}_3(\tau )+u(\tau -\tau_0)\int^\tau_{\tau_0}g(\tau
')\mathcal{A}_0(\tau ')d\tau ',\end{align} which are in some ways reminiscent of the Lorentz
transformation of the vector potential. One can extend these equations to a whole class of observers
along the lines discussed in the previous section; then, by imposing the gauge condition $A^\mu
_{\;\; ,\mu}=0$ and the wave equation $\Box A_\mu=0$ on the resulting equations, one would obtain the
nonlocal field equations for
$A_\mu (x)$. The main physical consequences of these nonlocal
equations and the nonlocal Maxwell equations of the previous section
are expected to be essentially the
same~\cite{17}, though there is no simple
(i.e. local) connection between
$\mathcal{A}_\mu$ and $\mathcal{F}_{\mu \nu}$.

\section{Discussion}\label{s6}

We have shown how to obtain nonlocal Maxwell's equations in the case of the electrodynamics of
accelerated systems. The resulting field equations in a global inertial reference system are Lorentz
invariant, since they are based on the definition of $\mathcal{F}_{\mu\nu}$ in equation~\eqref{eq9} and
the manifestly Lorentz-invariant nonlocal ansatz~\eqref{eq5} for
$\hat{\mathcal{F}}_{\alpha\beta}$. The nonlocal field equations may be expressed in terms of curvilinear
coordinates using the invariance of the 2-form $\mathcal{F}_{\mu\nu}dx^\mu \wedge dx ^\nu$.

The case of linearly accelerated observers has been treated explicitly for the derivation of nonlocal
Maxwell's equations. An interesting consequence of these equations is that nonlocality survives even
after the acceleration has been turned off. It would be interesting to subject this memory of past
acceleration to experimental test.

\renewcommand{\theequation}{\Alph{section}\arabic{equation}}
\appendix
\renewcommand{\thesection}{Appendix \Alph{section}}
\section{Resolvent kernel}\label{aA}
\setcounter{equation}{0}

The main integral relationship of the nonlocal electrodynamics of accelerated systems is of the general
form
\begin{equation}\label{eqA1} \phi (x)=\psi (x)+\epsilon \int^x_aK(x,y)\phi (y)dy,\end{equation} where
$\psi =\hat{\mathcal{F}}$, $\phi =\hat{F}$, $\epsilon =-1$, $a=\tau_0$ and $K(x,y)=k(y)$. We are
interested in the resolvent kernel $R(x,y)$ such that
\begin{equation}\label{eqA2} \psi (x)=\phi (x)+\epsilon \int^x_aR(x,y)\psi (y)dy.\end{equation} This
definition illustrates the complete reciprocity between $K$ and $R$, i.e. each is the resolvent of the
other. The expression for the resolvent kernel in terms of iterated kernels is readily available in the
standard sources on the Volterra integral equations~\cite{7,5,8}. However, care must be exercised as we
deal with systems of integral equations in~\eqref{eqA1} and~\eqref{eqA2}; in fact, in the case under
consideration in Section~\ref{s4}, $K$ and $R$ are $6\times 6$ matrices. The matrices involved in the
construction of iterated kernels do not commute in general; therefore, the order of the terms is important
here. Nevertheless, as described in this appendix, it turns out that the basic results of the theory
apply for
$N\times N$ matrices as well.

Following the method of successive approximation for solving~\eqref{eqA1} outlined in a previous work
(see~\cite{18}, Appendix A), we define the iterated kernels of $K$ via
\begin{align}\label{eqA3} K_1(x,y)&=K(x,y)\\
\intertext{and}
\label{eqA4} K_{n+1} (x,y)&=\int^x_yK(x,z)K_n(z,y)dz,\end{align} for $n=0,1,2,\ldots $. These emerge in
the successive approximation approach in the evaluation of double integrals of the form
\begin{equation}\begin{split}\label{eqA5} &\int^x_aK(x,y)\left[ \int^y_aK_n(y,z)\psi (z)dz\right]dy=\\
&\quad \int^x_a\left[ \int^x_zK(x,y)K_n(y,z)dy\right]\psi (z)dz\end{split}\end{equation} by changing the
order of integration in the triangular domain in the $(y,z)$-plane defined by the vertices $(a,a)$, $(x,a)$
and $(x,x)$. To interpret~\eqref{eqA5} properly, one must consider a summation over the appropriate
elements of the matrices involved in~\eqref{eqA5}. The resolvent kernel $R(x,y)$ is then given in terms of
the infinite series~\cite{18}
\begin{equation}\label{eqA6} R(x,y)=-\sum^\infty_{n=1} \epsilon^{n-1}K_n(x,y).\end{equation} This
expression may be written as
\begin{equation}\label{eqA7} R(x,y)+K(x,y)=-\epsilon
\sum^\infty_{n=1}\epsilon^{n-1}K_{n+1}(x,y)\end{equation} by adding $K(x,y)$ to both sides
of~\eqref{eqA6} and using~\eqref{eqA3}. Employing~\eqref{eqA4}, equation~\eqref{eqA7} is equivalent to
\begin{equation}\label{eqA8} R(x,y)+K(x,y)=-\epsilon \sum^\infty_{n=1}\epsilon
^{n-1}\int^x_yK(x,z)K_n(z,y)dz.\end{equation} Assuming that the operations of summation and integration
can be interchanged in~\eqref{eqA8}, which is the case under the
conditions of the Volterra-Tricomi uniqueness theorem, we find using~\eqref{eqA6} the following integral equation for $R$
\begin{equation}\label{eqA9} R(x,y)+K(x,y)=\epsilon \int^x_yK(x,z)R(z,y)dz.\end{equation} Starting
with~\eqref{eqA2} and thinking of $K$ as the resolvent of $R$, one can repeat the above analysis and
arrive at
\begin{equation}\label{eqA10} K(x,y)+R(x,y)=\epsilon \int^x_yR(x,z)K(z,y)dz.\end{equation} The integral
equations~\eqref{eqA9} and~\eqref{eqA10} illustrate the reciprocal relationship between $K$ and $R$.

It is interesting to note that 
\begin{equation}\label{eqA11} K_n(x,z)=\int^x_zK_r(x,y)K_{n-r}(y,z)dy,\quad r=1,2,\ldots
,n-1.\end{equation} This relation can be proved by induction. As an illustration, consider the iterated
kernel $K_{n+2}$ given by~\eqref{eqA4} as
\begin{equation}\label{eqA12} K_{n+2}(x,z)=\int^x_zK(x,y)K_{n+1}(y,z)dy.\end{equation} Now
using~\eqref{eqA4} again, we may express this relation as
\begin{equation}\begin{split}\label{eqA13} &\int^x_zK(x,y)\left[\int^y_zK(y,u)K_n(u,z)du\right]dy=\\
&\quad \int^x_z\left[ \int^x_uK(x,y)K(y,u)dy\right] K_n(u,z)du,\end{split}\end{equation} where we have
changed the order of integration over the triangular region in the $(y,u)$-plane defined by the vertices
$(z,z)$, $(x,z)$ and $(x,x)$. It follows from~\eqref{eqA13} that
\begin{equation}\label{eqA14}
K_{n+2}(x,z)=\int^x_zK_2(x,u)K_n(u,z)du,\end{equation} and so on. We
note that \eqref{eqA10} is a simple consequence of \eqref{eqA8} and
\eqref{eqA11}. 

Finally, let us consider the kernel
\begin{equation}\label{eqA15} \tilde{K}(x,y)=\Lambda^{-1}(x)K(x,y)\Lambda (y)\end{equation} corresponding
to~\eqref{eqA1} for $\psi=\mathcal{F}$ and $\phi =F$. It is simple to check that the resolvent kernel in
this case is 
\begin{equation}\label{eqA16} \tilde{R}(x,y)=\Lambda^{-1}(x)R(x,y)\Lambda (y).\end{equation} It follows
that $\tilde{r}(\tau ,\tau ')$ in equation~\eqref{eq13} is the resolvent kernel of
\begin{equation}\label{eqA17} \tilde{K}(\tau ,\tau ')=-\Lambda^{-1}(\tau )\frac{d\Lambda (\tau
')}{d\tau'}\end{equation} using equations~\eqref{eq8} and~\eqref{eqA15}.

\section{A simple example}\label{aB}\setcounter{equation}{0}

In this appendix, we eschew matrix-valued kernels and consider a simple case with
\begin{equation}\label{eqB1} K(x,y)=\alpha (x)\eta (y),\end{equation} where $\alpha$ and $\eta $ are
smooth real functions. Let us define the function
\begin{equation}\label{eqB2} G(u)=\int^u_aK(z,z)dz.\end{equation} Then, it is straightforward to show
that the iterated kernels are given in this case by
\begin{equation}\label{eqB3} K_{n+1}(x,y)=\frac{1}{n!}[G(x)-G(y)]^nK(x,y),\end{equation} for
$n=0,1,2,\ldots $. It follows from~\eqref{eqA6} that the resolvent kernel is 
\begin{equation}\label{eqB4} R(x,y)=-K(x,y)\exp \left( \epsilon \int^x_y K(z,z)dz\right).\end{equation}
Unfortunately, this simple result cannot be immediately extended to matrix-valued kernels, since the
matrices involved in the evaluation of iterated kernels do not commute in general. This is the case even
when $\alpha (x)$ is the unit matrix.

\section{Uniform acceleration}\label{aC}\setcounter{equation}{0}

Let us consider the kernel $K(\tau ,\tau ')=k(\tau ')$ given by equation~\eqref{eq8}. It does not appear
possible in general to obtain an expression for the resolvent kernel $r(\tau ,\tau ')$ in closed form. We
therefore consider the special case of uniform acceleration where $k$ is a constant matrix. In this case,
\eqref{eq8} can be written as
\begin{equation}\label{eqC1} \frac{d\Lambda}{d\tau }=-k\Lambda ,\end{equation} which has the solution
\begin{equation}\label{eqC2} \Lambda (\tau )=e^{-k\tau}\Lambda (0).\end{equation} For a constant $k$, the
iterated kernels are given by
\begin{equation}\label{eqC3} K_{n+1}(\tau ,\tau ')=\frac{1}{n!}(\tau -\tau')^nk^{n+1},\end{equation} for
$n=0,1,2,\ldots $. It follows that the resolvent kernel is of the convolution type and is given by
$(\epsilon =-1)$
\begin{equation}\label{eqC4} r(\tau ,\tau ')=-k\;\; e^{-k(\tau -\tau')}.\end{equation} Furthermore, using
equation~\eqref{eq13} we find that
\begin{equation}\label{eqC5} \tilde{r} (\tau ,\tau ')=-\Lambda ^{-1}(0)\; k\; \Lambda (0).\end{equation}

More explicitly, for an observer that accelerates uniformly from rest at $\tau =\tau _0$ along the
positive $z$-direction with acceleration $g$, $\Lambda$ is given by~\eqref{eq16} with $C=\cosh g(\tau
-\tau _0)$ and $S=\sinh g(\tau -\tau _0)$. It follows that $\Lambda (\tau_0)$ is the unit matrix in this case
and~\cite{18}
\begin{equation}\label{eqC6} k=\begin{bmatrix} 0 & -{\bf g}\cdot {\bf I}\\ {\bf g}\cdot {\bf I} &
0\end{bmatrix},\end{equation} where ${\bf g}\cdot {\bf I}=gI_3$. Thus $\tilde{r}(\tau ,\tau ')=-k$ is a
constant matrix for uniform linear acceleration.

In the case of uniform rotation with frequence $\Omega_0$ about the $z$-axis on a circle of radius $r_0$,
$\Lambda$ is of the form given in~\eqref{eq16} with~\cite{18}
\begin{equation}\label{eqC7} U=\begin{bmatrix} \gamma \cos \varphi & \gamma \sin \varphi & 0\\ -\sin
\varphi & \cos \varphi & 0\\ 0 & 0 &\gamma \end{bmatrix},\quad V=\beta \gamma \begin{bmatrix} 0 & 0 & 1\\
0 & 0 & 0\\ -\cos \varphi & -\sin \varphi & 0\end{bmatrix},\end{equation} where $\beta =r_0\Omega_0$,
$\gamma=(1-\beta^2)^{-1/2}$ and $\varphi =\gamma \Omega_0(\tau -\tau _0)$. In this case~\cite{18}
\begin{equation}\label{eqC8} k=\begin{bmatrix} \text{\boldmath$\Omega$\unboldmath} \cdot {\bf I} & -{\bf
g}\cdot {\bf I}\\
 {\bf g}\cdot {\bf I} & \text{\boldmath$\Omega$\unboldmath} \cdot {\bf I}\end{bmatrix},\end{equation}
where \boldmath$\Omega$\unboldmath\; is along the axis of rotation with $\Omega_3=\gamma^2\Omega_0$ and
${\bf g}$ is the centripetal acceleration with its only nonzero component given by $g_1=-\beta
\gamma^2\Omega_0$. One can show that $\tilde{r}$ is constant,
\begin{equation}\label{eqC9} \tilde{r}=-\gamma \Omega_0\begin{bmatrix} I_3 & 0 \\ 0 &
I_3\end{bmatrix},\end{equation} for uniform rotation about the $z$-axis in accordance with~\eqref{eqC5}.

\section{Linear acceleration}\label{aD}\setcounter{equation}{0}

The purpose of this appendix is to find the resolvent kernel associated with
\begin{equation}\label{eqD1} K(x,y)=k(y)=-g(y)\begin{bmatrix} 0 & I_3\\ -I_3 &
0\end{bmatrix}\end{equation} for a linearly accelerated observer. It follows from~\eqref{eqA3}
and~\eqref{eqA4} by straightforward integration using~\eqref{eq15} that the iterated kernels are given by
\begin{equation}\label{eqD2} K_{n+1}(x,y)=\frac{(-1)^{n+1}}{n!} g(y)\sigma^{n} (x,y)M_{n+1},\end{equation}
where $n=0,1,2,\ldots $, and $\sigma (x,y)$ is given by
\begin{equation}\label{eqD3} \sigma (x,y):=\theta (x)-\theta (y)=\int^x_yg(z)dz.\end{equation}
 Moreover, $M_{n+1}$ are $6\times 6$ matrices defined by
\begin{equation}\label{eqD4} M_{2n+1}:=\begin{bmatrix}0 & I_3\\ -I_3 & 0\end{bmatrix} ,\quad
M_{2n+2}:=\begin{bmatrix}J_3 & 0\\ 0 & J_3\end{bmatrix},\end{equation} where $(J_k)_{ij}=\delta
_{ij}-\delta _{ik}\delta_{jk}$, so that $J_3=-I^{\;\; 2}_3$. Thus the series~\eqref{eqA6} can be summed in
this case and the result is
\begin{equation}\label{eqD5} r(x,y)=g(y)\begin{bmatrix} \mathcal{S}J_3 & \mathcal{C}I_3\\ -\mathcal{C}I_3
&
\mathcal{S}J_3\end{bmatrix},\end{equation} where $\mathcal{C}$ and $\mathcal{S}$ are given by
\begin{equation}\label{eqD6} \mathcal{C}=\cosh \sigma (x,y),\quad \mathcal{S}=\sinh \sigma
(x,y).\end{equation}

We are also interested in $\tilde{r}(x,y)$ given by~\eqref{eq13}. It follows from a straightforward
calculation that
\begin{equation}\label{eqD7} \tilde{r}(x,y)=g(y)\begin{bmatrix} 0 & I_3\\ -I_3 &
0\end{bmatrix}\end{equation} using~\eqref{eq16}. Thus we find that $\tilde{r}(x,y)=-k(y)$ in agreement
with~\eqref{eq18}.

\end{document}